\documentclass[11pt,a4paper]{article}

\usepackage[english]{babel}
\usepackage[numbers]{natbib}
\usepackage{graphicx}
\usepackage{amsmath,amssymb,amsfonts}
\usepackage{mathrsfs}
\usepackage{manyfoot}
\usepackage{booktabs}
\usepackage{multirow}
\usepackage{csquotes}
\usepackage{xurl}
\usepackage{rotating}

\usepackage[
    colorlinks=true,
    linkcolor=black,
    citecolor=black,
    urlcolor=blue,
    pdftitle={The Illusion of an Automated German Census: Assessing the Feasibility of the 2031 Register-Based Census via Simulation},
    pdfauthor={Rainer Schnell and Severin V. Weiand},
    pdfsubject={Register-based Census, Census Methodology},
    pdfkeywords={register census, record linkage, data quality, simulation, quasi-identifiers, administrative data, undercount, undercoverage},
    pdfcreator={LaTeX},
    pdfdisplaydoctitle=true
    ]{hyperref}
\title{The Illusion of an Automated German Census: Assessing the Feasibility of the 2031 Register-Based Census via Simulation}

\author{
    Rainer Schnell\thanks{Corresponding author: rainer.schnell@uni-due.de}
    \and Severin V. Weiand\\
    \parbox{\textwidth}{
        \centering
        \small Research Methodology Group\\
        \small University of Duisburg-Essen, Germany
    }
 }

\date{\today}
\begin{document}

\maketitle

\begin{abstract}
The next German Census (2031) is planned to be the country's first fully register-based census, relying exclusively on administrative data sources. In addition to the population registers, ten further administrative registers are envisaged, covering a wide range of information, from driving licences to agricultural social insurance. This study simulates potential data-quality challenges in linking these registers using quasi-identifiers (QIDs), such as names and dates of birth. Using a synthetic population dataset (N=80\,206\,770), we systematically vary experimental factors related to register characteristics. Across the 1\,458 experimental conditions studied, results vary considerably. Some plausible scenarios range from 500\,000 to 1\,000\,000 false-negative linkage errors per linked register. These findings suggest that implementing an automated register-based census using the currently envisaged registers may prove challenging and could require substantial manual post-processing to achieve acceptable data quality.

\end{abstract}

\section{Introduction}
Like many national statistical offices in Europe, the Federal Statistical Office (Destatis) is currently planning to transition from a register-assisted census to a fully register-based census~\cite{KoernerKrauseRamsauer2019, SoellnerKoerner2022, BMI2026Registerzensus}. Under this approach, sample-based population surveys -- conducted in the 2011 and 2022 censuses -- would be discontinued. Except for the census of buildings, all census data would be derived from administrative and statistical registers.

 As the 2031 Census will employ a new methodology for determining population counts, rigorous methodological evaluations before its implementation are particularly important. Such studies could help identify potential shortcomings at an early stage to prevent biased estimates. However, no methodological studies evaluating the new approach have been published to date. The simulation presented in this paper aims to address this gap by independently investigating potential challenges associated with the forthcoming census.

\section{Register Census in Germany}

In a register census, the census relies exclusively on existing administrative and statistical registers. Since Germany has no centralised population register, the core component of the census will be the combined resident register, derived by merging and deduplicating the decentralised resident registers of individual municipalities. A further key requirement of a register-based census is the existence of appropriate registers with sufficient data quality that can be used to enhance the data stored in the resident registers.

To this end, the Register Modernisation Act was enacted in 2021. In particular, under the Identification Number Act of 2021, the tax number will be introduced as a unique ID (UID) in about 50 existing registers. The tax number serves as an interim solution and will later be replaced by a standalone ID. The census model aims to use the UID to link the registers in the census. However, requiring the UID in all registers until 2031 may be ambitious. Therefore, at least some registers will likely need to be linked using record-linkage techniques. These linkages will have to rely on personal characteristics, known as quasi-identifiers (QIDs), such as names and dates of birth. Little is known about the data quality of QIDs in administrative data.\footnote{Additionally, the tax number itself is added to a register by an identification request to the federal tax office. This identification request is also a record-linkage problem, since the records in the tax register must be linked to records in other registers based on the available QIDs. The introduction of the UID is therefore not a general solution, but rather a shift of the problem to a different record-linkage process. It will probably take decades for registers to be filled with UIDs generated through the registration process, not the identification request.}

\subsection{Census Modules}

The register census will comprise five modules~\cite{SoellnerKoerner2022}:  housing~\cite{KrauseZimmermannHerda2022,HennigGebhard2023}, households, education~\cite{GrimmHerzogRheiner2022}, labour~\cite{BuffGansererRamsauer2024} and, most important, population. 

The population module will be based on the combined resident register of the municipalities. The current draft of the Census Act~\cite{BMI2026Registerzensus} does not explicitly define the population. Still, it seems to rely on the same definition of the \enquote{usual place of residence} as the Census 2011 and 2022.\footnote{In the Census 2011 and the Census 2022, Germany used a definition based on the place where the person should be registered, if the person is acting according to law. According to the UNECE survey~\cite[p. 66]{UNECE2026}, only a few of the 45 reporting countries used similar definitions. 31 countries used a \enquote{usually resident} population definition, and 10 others used a \enquote{registered population} definition.} The Conference of European Statisticians~\cite{UNECE2025} recommends that, for the 2030 censuses, the population be defined by usual residence as the place where the enumerated person usually spends their daily rest. Continuing the use of the previous census definitions will make European comparisons difficult and will require no less complex record-linkage operations than the Census 2022.

Each other module faces specific challenges, primarily due to the lack of dedicated administrative registers. This is particularly evident in the housing module. Since no suitable register is expected to be available  by 2031, Destatis has already decided to conduct a full population survey for this module~\cite[p. 3]{BundesministeriumsdesInnern2025}. 

The main problem in the household module is that resident registers do not contain household identifiers. Therefore, households have to be generated. This process is error-prone, when no survey data is available~\cite{BaumstarkFuernrohrReichert2026}.

The education and labour modules suffer from a lack of appropriate data content, since no educational register and no population-covering labour data are available. Hence, it is planned that these mainly rely on models~\cite{BuffGansererRamsauer2024, GrimmHerzogRheiner2022}. Destatis has not revealed how these models are constructed, how and whether they are validated, or which data will be linked and used.

\subsection{Sign-of-Life Approach}

To identify coverage errors in the resident registers, the sign-of-life approach will be used~\cite{SoellnerKoerner2022}. In this approach, the resident registered population is linked to administrative contacts in other administrative registers (\textit{comparison registers}). Currently, 10 registers are planned to be utilised as comparison registers~\cite[p. 108--110]{BMI2026Registerzensus}. 

Mentioning a QID in this law does not guarantee that the QID attributes are available for the whole population covered in the register. No comparative publication reports on data quality for these registers, including the amount of missing data (neither by design nor by nonresponse). Hence, the available evidence on data quality is sparse. However, some details are known. For example, although the register of foreign nationals can store addresses, it generally contains pointers to the responsible municipal authority~\cite[p. 12]{BundesamtfurMigrationundFluchtlinge2024}. In an official evaluation of this register, the address information is declared as throughout not up to date~\cite[p. 19]{DeutscherBundestag2020}.

\begin{figure}
    \centering
    \includegraphics[width=1\linewidth]{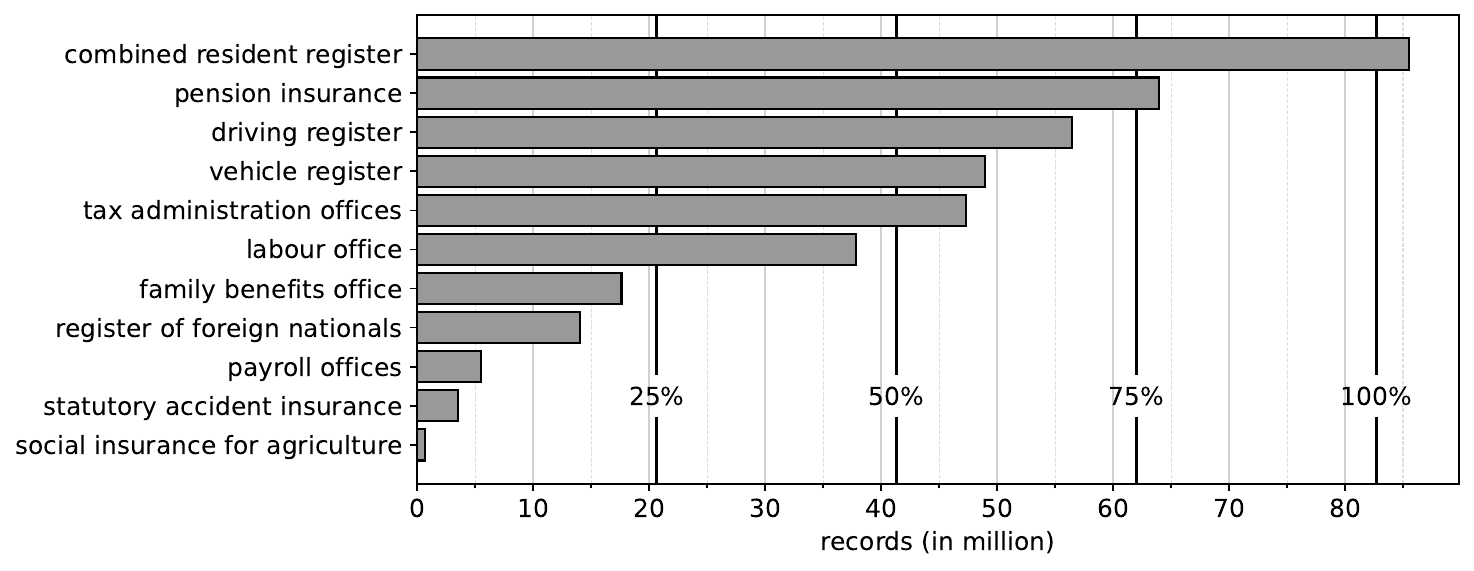}
    \caption{Estimated sizes of individual records in the registers. Data are taken mostly from the registers' annual reports. Black vertical lines indicate proportions of the 2022 census population for reference.}
    \label{fig:sizes}
\end{figure}

There are almost no published numbers on the number of records in the registers in consideration for the register census. Figure \ref{fig:sizes} shows our estimated counts based on annual reports. The size of registers varies between 675\,000 and 64 million records~\cite{Kremer2026}. By comparison, the combined resident register in the 2022 census contained 85\,542\,382 records~\cite{GrobeckerSchmollinger2025}. Therefore, it can be estimated, that the comparison registers cover between 1\% and 77\% of the census population. 

In the sign-of-life approach, the number of signals generated each year is of central importance. However, there is hardly any information on these distributions. It can be expected that registers with recurring payments will be among the most useful for the register census. It should be noted that most registers do not cover a random sample of the population; therefore, some subpopulations are covered by few registers. For example, among the considered registers, children are covered only by family benefits. Hence, the effect of linkage errors on the detected signs of life will differ between subpopulations, resulting in an inevitable linkage bias.

According to Destatis, a detected life-sign will be accepted when it indicates an administrative contact within one year ahead of the census reference date~\cite{SoellnerKoerner2022}. There are no published plans to weight life signs or create a residency index, as is done in Estonia~\cite{Tiit2017}.

Estonia, for example, uses 14 registers in its sign-of-life approach~\cite{StatisticsEstonia2021}. Compared to the registers under consideration, Estonian registers contain more frequent and geolocated life signs, such as digital prescriptions, medical bills, dental care, and legal proceedings. In contrast, the German driving licence register, for instance, does not include addresses and does not record fines, but primarily captures infrequent administrative events. Another example is the statutory accident insurance, which, despite its name, does not cover accidents but only business registration and deregistration.

\subsection{Residence Analysis}
When multiple addresses for one person are identified through the comparison register linkage, a residence analysis will clarify the correct address. Destatis estimates that about 10 million cases must be clarified in the residence analysis~\cite{WojcikSoellner2025}.\footnote{During the summer 2025, the residence analysis has been tested in a sample of 100\,000 cases. According to Destatis, the results will not be published before Spring 2027. However, due to the small sample and expected error rates in identifiers below 1\%~\cite{Weiand2024}, the number of linkage problems detected by this test will result in an underestimation of problems to be expected in the census.} In comparison, this is almost the same size as the household survey of the 2022 census (11\,531\,900 people;~\cite[p. 19]{KlinkLorentz2022}).

The proposed sign-of-life is of limited use for identifying coverage errors. For example, the usual place of residence is difficult to identify using the proposed registers. The current approach assumes that the comparison register contains the person's actual address. This might be the case for data-processing errors but not for intentional incorrect registrations, which probably accounts for the vast majority of coverage errors (e.g., students still registered at their parents' address). Hence, it seems debatable whether the residence analysis will identify an address different from those in the resident register. Destatis does not publicly discuss these aspects. However, the statistical problems of the sign-of-life approach are the topic of our ongoing research and will be covered in a separate paper.

\subsection{Previous Results on the Linkage of Registers in Germany}

Published information on the quality of the data used to link German administrative registers is limited. Destatis has not published any details on register quality after the 2022 Census.\footnote{The central publication here~\cite{Wittmaack2025} contains no quality metrics for either data quality or linkage results. The 2022 Census quality report~\cite{DestatisZensus2022QB2024} contains only the number of rejected records at data intake \enquote{less than one per cent}. With the total number of processed individual records~\cite{Aude2024} at approximately 94 million, this would correspond to about 940\,000 erroneous records.}

However, there are two relevant results on linking German administrative data. Both reports are difficult to find and have therefore not been cited in the literature.

From a technical perspective, linking multiple registers is similar to introducing the tax identification number after 2007. The identification problems only became apparent through a parliamentary inquiry~\cite{Bundestag2014}. On average, between 2007 and 2014, the available data could not uniquely assign 23.5\% of the generated entries. Subsequently, an additional approximately 106\,000 entries had to be deleted, as they were identified as duplicates. This corresponds to 0.113\% duplicates. No more recent figures on tax identification number assignments are available.

A recent empirical example of the problems of merging older data holdings with files containing a uniform identification number is the first pilot to introduce the tax ID into the national weapons register: only about 90\% of individuals could be assigned an ID~\cite{Bundesverwaltungsamt2024}. 
The publication does not include a linkage-bias assessment or an analysis of heterogeneity between local data providers. Therefore, the results could be considered optimistic.

\section{Methods}

To quantify the expected quality of a register-based census, census aspects relevant to linkage quality were simulated. The simulation begins by generating a base dataset representing the true population at the time of the census. Second, the combined population registers and comparison registers were derived from the base dataset. Factors affecting register quality were incorporated as experimental factors in this generation process. The resulting registers were then corrupted by introducing multiple types of errors. Finally, the registers were linked using record-linkage methods. All steps are described in more detail below. A step-by-step description of the data generation is available as technical report~\cite{WeiandSchnell2026}. 

\subsection{Base Dataset}

The base dataset draws primarily on the publicly available results of the 2011 Census and the 2011 Microcensus Scientific-Use-File. The Microcensus is a mandatory 1\% sample of the population, giving more demographics details than the census. Since obtaining additional data sources, such as name distributions or migration patterns, had already taken several years and acquiring new data would require several more years, 2011 was chosen as the reference year. For the purpose of the simulation, census results were considered true, disregarding methodological issues of the last censuses~\cite{SchnellLenzSchmidt2026}. 

The first step in creating the base dataset, which comprises individual-level records, was to generate households. A key priority was to preserve household composition, particularly the age and gender distributions within households, as well as their geographical distribution at the small-area level. The aim was to geolocate the dataset at the 100m grid-cell level, which represents the finest spatial resolution at which census results are published.

Previous approaches to spatial microsimulation~\cite{TemplMeindlKowarikDupriez2017,Kavroudakis2015, Whitworth2022,TantonEdwards2013} first generate a population dataset by expanding datasets similar to the Microcensus. The resulting datasets are then distributed across geographical units using optimisation methods such as simulated annealing, while taking additional constraints for those units into account. These approaches are limited to a few thousand geographical units, often even fewer than 100. In our simulations, these methods do not scale to the large number of geographical units required here (3.2 million grid cells). 

A new method was therefore developed to generate the base dataset. Unlike previous approaches, generation began at the level of geographical units. At first, the grid-cell level and then the municipal level were considered. Constraints on age, gender, nationality, household size, and family type were applied at these levels, initially using coarse categories. The households selected from the Microcensus were assigned to a flat in a building within a grid cell. Buildings were assigned nearby real-world addresses from OpenStreetMap. Because collective living quarters pose challenges for census operations, about 18\,000 corresponding units were simulated, containing about 747\,000 people.

In the next iteration, more detailed attribute values were assigned. First, for each person record exact nationality, country of birth, and place of birth were fixed. This process considers changes in identifiers due to historical events, such as the dissolution of the Soviet Union and Yugoslavia~\cite{WeiandAntoniFilser2026}. For people born in Germany, the geographical distance between their current place of residence and their place of birth was estimated conditional on their age and the size of their current municipality.

In the next step, for each person record the exact date of birth was determined, conditional on year, month, and day of the week~\cite{Lerchl2005}. Since linking multiple births is particularly challenging because they share many personal characteristics, multiple births were simulated in detail. Finally, the birth name, surname, and first name were fixed. For birth names and surnames, the kinship relationships among household members were considered. For example, children were assigned their parents’ married name as both their birth name and surname, provided that the parents were married. The married name was selected based on the distribution of naming choices and the naming law in force at the estimated time of the marriage. In particular, the number of name components (tokens) in first names was simulated according to the age within the person record. We are not aware of any linkage simulation considering conditional dependencies of QIDs at this level of detail.

The following attributes were simulated: name (forename, surname, birth name), full date of birth (day, month, year), sex (male, female), address (street name, house number, postcode, postcode area name, official regional code, 100m-grid-cell ID, special area flag), place of birth, country of birth, nationality, household relationships, family type, person ID (UID), household ID, building ID, and multiple-birth ID.

\clearpage

\subsection{Assumptions and Experimental Design}
\begin{table}[t]
    \centering
    \begin{tabular}{lp{8cm}}
        \toprule
        \textbf{QID set} & \textbf{attributes} \\
        \midrule
        core & forename, surname, full date of birth, sex \\
        core+PoB & forename, surname, full date of birth, sex, place of birth \\
        all & forename, surname, birth name, full date of birth, sex, place of birth, country of birth, 100m-grid ID \\
        \bottomrule
    \end{tabular}
    \caption{QID sets used in the simulation.}
    \label{tab:QIDsets}
\end{table}

For the simulation, three sets of QIDs were used (see Table \ref{tab:QIDsets}). The first QID set is based on the census core attributes from the 2022 census~\cite{StatistischesBundesamt2024d}. In the second QID set, the core attributes are extended by the place of birth~\cite{Schnell2022a}. The third set includes all potentially available and useful QIDs for the linkage. Country of birth, rather than citizenship, was used because it changes less often. Since the grid ID usually captures all address information useful for linking, it was preferred over other QIDs~\cite{Weiand2024}. In this simulation, academic title and marital status were not considered, as these attributes are missing in most registers by law. 

It is reasonable to assume that most duplicates can be identified during the deduplication process of the combined resident register, since additional attributes (e.g., names of children and the marriage partner, or former residences) can be used~\cite{Wittmaack2025}. In the 2022 census and the 2011 census tests, the overall number of duplicates was between 500\,000 and 700\,000~\cite{Wittmaack2025, Braun2004}. It is expected that the number of duplicates will decrease over time, since the decentralised registers started using an automated data exchange of resident register entries~\cite{Wittmaack2025}. The effect of undetected duplicates in these registers is therefore not considered in the simulation.

The most important source of errors in the resident registers seems to be under- and overcoverage, as the results of the 2022 census demonstrated~\cite{SchnellLenzSchmidt2026}. Undercoverage mainly occurs when people live in a municipality but are not registered there. Examples include students still registered at their parents' home or unregistered migrants. The latter population is not covered in the previous censuses. By definition, it cannot be estimated using a resident census, since these people will not appear in any of the comparison registers. The latest estimates range between 180\,000 to 520\,000 people~\cite{Vogel2016}. Although the UNECE~\cite{UNECE2025} recommends that European censuses include undocumented migrants, their inclusion can be considered a legal problem regarding the definition of the population. In the simulation described, unregistered migrants are not modelled. 
\clearpage
Overcoverage errors are mostly due to people who are registered in a municipality but no longer live there. For the most part, these seem to be emigrants, or people moving to a different municipality but did not deregister. This population also contributes to undercoverage in the target municipality. Especially emigrating foreigners, who did not deregister are reported to be a major source of error~\cite{Destatis2024QBFortschreibung}.

\subsection{Experimental factors in the comparison registers}

\begin{table}[t]
    \centering
    \begin{tabular}{ll}
        \toprule
        experimental factor & levels \\
        \midrule
        identifiers used (QID set)& all, core+PoB, core\\
        address outdating probability& 0.02, 0.04, 0.06\\
        duplicates in the register& 0, 0.01\\
        proportion of updated records& 0, 0.5, 1.0\\
        overlap of register to base data& 0.25, 0.5, 0.75\\
        proportion of records with UIDs& 0, 0.7, 0.9\\
        error factor& 4, 8, 16\\
        \bottomrule
    \end{tabular}
    \caption{Experimental factors and their levels.}
    \label{tab:experimentalfactors}
\end{table}

The base dataset is used to derive the combined resident register. To simulate coverage errors, the net divergence results between the 2022 census and the 2022 population projection of official statistics were used. Gross divergence was simulated by the probability of an address being outdated (2\%, 4\%, 6\%), corresponding to the probability of residential change.\footnote{Results from the follow-up  survey of the census) of the 2011 Census in Hamburg showed discrepancies of addresses around 4\%~\cite[p. 36]{StatistikamtNord2015}. A report on the 2011 census test ~\cite[p. 8]{Braun2004} also noted that 4\% of households receiving social security benefits had an incorrect address.}
First, 75\% of address errors were added by the same function already used to determine the place of birth. This process creates coverage errors for all municipalities. Second, the remaining 25\% were distributed without distance limits. Finally, a synthetic overcoverage population was generated and added to all municipalities with lacking  overcoverage observations. 

As noted above, a critical assumption of the residence analysis proposed by Destatis is that the current true address is present in at least one of the relevant comparison registers. Hence, the comparison register records were drawn from the base dataset. Compared to the resident registers, the comparison registers contain fewer QIDs (usually just names, date of birth, gender, and address. Therefore, the comparison registers might contain duplicates. To estimate the effects of duplicates in the comparison registers, duplicate rates of 0\% and 1\% were simulated~\cite{SedmihradskyEisenmengerBurmeister2012}.
\clearpage
An outdating process was implemented to account for the ageing of register information.  The process used an update probability (0\%, 50\%, 100\%), where 0\% indicates that records are entered and never updated (e.g., in an educational register), and 100\% implies that the information in the register is up-to-date for all records (e.g., a tax register). The outdating process consisted of sub-processes for naturalisation, change of sex, renaming of country of birth, renaming of place of birth, change of residence, and marriage.

From the resulting outdated dataset, subsets of 25\%, 50\% and 75\% were sampled using simple random sampling. These samples represent the set of records in the registers  relevant to the census. As Figure \ref{fig:sizes} shows, the resulting dataset sizes correspond to the range of the comparison registers. Since most registers do not contain children, only records representing people 18 years or older were selected.

As the registers are intended to be linked using the UID, the proportion of records with a UID was simulated (0, 0.7, 0.9). The UID was assumed to be always correct. Hence, only records without a UID have to be linked using record linkage techniques.  As noted, the German UID is generated by linking the comparison register to the tax register. Because this process is also prone to linkage errors, records without a UID were sampled based on the number of data errors in the record.

Both the resident and comparison registers were corrupted using a data-corruption function to account for data-entry errors, such as typing errors. 
There is almost no published information on the data quality of QIDs in administrative registers. Therefore, the error rates are based on the only available study~\cite{Weiand2024}, which examined the resident register of all young people visiting schools in Hamburg. As error rates differed between people born in Germany and abroad, separate levels of data corruption were applied.

Since the error rates in the cited study refer to observed changes over a two-year period, to account for cumulative effects over time, they were multiplied by an error factor (4, 8, 16). Both the resident and comparison registers were always affected by the same error factor, but with different random seeds. 

The error factor was varied between municipalities using a hard-to-link index~\cite{Schnell2019exp}.\footnote{Since the hard-to-link index $htl$ were integers in the range 0-4, the error factor was multiplied by  $  0.75 + \text{htl}/8$.} The index accounts for the varying quality of the resident registers in the municipalities and was developed by predicting the deviation between the 2011 census results and the population projection using various population-related properties of the municipality, such as homeownership rate and school dropout rate (see~\cite{WeiandSchnell2026}). Furthermore, people living in collective living quarters were assigned a higher error rate.

Finally, the Hamburg study indicated that token count is the most common source of error when comparing names. These errors occur when names have multiple tokens (e.g., the forename Thomas Peter) but only a subset of tokens is present in the register. Most commonly, this subset consists only of the first token (e.g., Thomas), where the resident register usually contained the full name and the school data contained only the first token. This error was also found in the 2011 census~\cite{SedmihradskyEisenmengerBurmeister2012}. Although the resident registers probably contain the full name, the comparison registers likely do not. Thus, the names were shortened to the first token of the forename (first name), surname (last name), and birth name. This shortening increases linkage quality, since similarities among shortened names proved higher than those among full names.

\subsection{Linkage Methods}

For probabilistic record linkage, we used Splink~\cite{Linacre2022}, as does Destatis~\cite{Kremer2026}.\footnote{Python 3.10.12 and Splink 5.0.0 with DuckDB 1.5.6 was used. Simulation and linkage was run on an Ubuntu 22.04 machine with an AMD EPYC 7713 CPU (64 cores, 128 threads, max. 3.67 GHz), and 1 TB RAM.} 
Splink implements a variant of the Fellegi-Sunter model~\cite{Fellegi1969},  which still generally shows to be the best record linkage method in comparisons~\cite{GomatamCarterAriet2002,Campbell2009,Bohensky2016,Weiand2022}.

Jaro-Winkler similarity was used for all names (thresholds at 0.9 and 0.7) and place of birth (threshold at 0.9)~\cite{Christen2012}. Exact agreement was used for all other QIDs. As is customary in German official statistics, the linkage was designed to minimise false-positive results~\cite{Schnell2021}. 

The combination of the experimental factors (see Table \ref{tab:experimentalfactors}) resulted in 1\,458 individual linkage scenarios. To speed up computations, the estimation was split into two steps. First, initial estimates were generated using a population sample for a subset of scenarios. Second, the parameters were refined for all scenarios based on the full dataset. 

For training, the data were blocked on first name, last name, full date of birth, and 100m-grid ID. For initial estimates, the total population of a federal state with about 1 million people was used. In the second step, the estimates of the probabilities that QIDs agree at random and the proportion of matches were refined. Every scenario has its own refined set of parameters.

The linkage was implemented in a two-stage process. First, exact matching was performed on the census core QIDs, after which Splink was applied exclusively to the subset of records that remained unmatched. In the second stage, records were partitioned using eight distinct, non-hierarchical blocking schemes based on subsets of the available QIDs. Records were compared if they matched on at least one blocking scheme. A blocking scheme was included only if all constituent QIDs were present in the QID set. Links were afterwards resolved using one-to-one (no duplicates) or one-to-many (with duplicates) linkage constraints.

\section{Results}
Linkage quality is measured using precision (positive predictive value) and recall (sensitivity)~\cite{Christen2012}. With 1\,458 linkage scenarios, assessing the effects of individual experimental factors is not straightforward. The results are therefore summarised using several complementary perspectives.

\begin{table}[t]
    \centering
    \begin{tabular}{llrr}
    \toprule
    error factor & QID set & precision (in \%) & recall (in \%) \\
    \midrule
    \multirow[t]{3}{*}{4} & all & 99.99 & 99.31 \\
     & core+PoB & 99.98 & 99.13 \\
     & core & 99.98 & 96.19 \\[2mm]
    
    \multirow[t]{3}{*}{8} & all & 99.98 & 98.90 \\
     & core+PoB & 99.97 & 98.45 \\
     & core & 99.98 & 94.31 \\[2mm]
    
    \multirow[t]{3}{*}{16} & all & 99.93 & 97.27 \\
     & core+PoB & 99.95 & 96.79 \\
     & core & 99.96 & 90.82 \\
    \bottomrule
    \end{tabular}
    \caption{Mean precision and recall for all scenarios with 0\% UID fill and 0\% duplicates.}
    \label{tab:meanpr}
\end{table}

Table \ref{tab:meanpr} shows the average precision and recall for all scenarios, broken down by error factor and QID set. Overall, precision is very high, and recall is moderately high. Recall decreases with lower error factors. When all QIDs are available, recall is highest, closely followed by the core identifiers combined with place of birth. If only core identifiers are used, recall is noticeably lower.

Precision and recall can be misleading when linking large datasets, as a high number of true hits can obscure the absolute magnitude of false positives and false negatives. In a census, large absolute numbers of false negatives are particularly important, as they necessitate additional work to resolve missing links.

Therefore, Table \ref{tab:FPFN} shows the absolute numbers of false positives and false negatives for the best-case and worst-case register scenarios. The worst case is a register with 0\% UID fill, 1\% duplicates, and 0\% update probability; the best case, by contrast, is a register with 90\% UID fill, 0\% duplicates, and 100\% update probability. The table shows that even in the worst-case scenario with the highest error rate, the number of false positives is low. Under high error rates, even in the best-case scenario, false negatives become problematic, especially when only the core identifiers are available. Thus, the highest number of observed false negatives in the best-case scenario (error factor 16, core identifiers) is 731\,874 cases. 
\clearpage
The worst-case scenario shows the same patterns. Again, the core identifiers cause many false negatives, even with low error rates. However, in almost all cases, the number of false negatives exceeds 1 million.

\begin{table}[t]
    \centering
    \begin{tabular}{llrrrr}
    \toprule
    && \multicolumn{2}{c}{worst-case} & \multicolumn{2}{c}{best-case}\\\cmidrule(lr){3-4}\cmidrule(lr){5-6}
    error factor & QID set & FP & FN & FP & FN\\
    \midrule
    \multirow[t]{3}{*}{4} & all & 12,070 & 960,675 & 467 & 53,439 \\
     & core+PoB & 19,623 & 1,084,397 & 279 & 53,413 \\
     & core & 21,798 & 2,764,350 & 261 & 135,735 \\[2mm]
    \multirow[t]{3}{*}{8} & all & 22,291 & 1,212,247 & 2,341 & 120,945 \\
     & core+PoB & 25,132 & 1,397,614 & 2,685 & 162,268 \\
     & core & 27,914 & 3,515,691 & 1,310 & 450,630 \\[2mm]
    \multirow[t]{3}{*}{16} & all & 61,005 & 1,946,974 & 2,350 & 365,196 \\
     & core+PoB & 38,549 & 2,093,670 & 3,319 & 426,243 \\
     & core & 38,627 & 4,915,298 & 1,378 & 731,874 \\
    \bottomrule
    \end{tabular}
    \caption{Number of false positive (FP) and false negatives (FN) for a worst-case (0\% UID fill, 1\% duplicates, 0\% updates) and best-case scenario (90\% UID fill, 0\% duplicates, 100\% updates). Both scenarios assume 4\% address outdating and 50\% overlap.}
    \label{tab:FPFN}
\end{table}

Finally, the 1\,458 scenarios are summarised by three linear regression models.\footnote{The results of the linear model with a log-transformed dependent variable do not differ qualitatively from those of a Poisson regression. Stata 19 was used for computing the regressions.} For each set of QIDs, the logarithm of the number of false negatives is predicted with all experimental factors coded as dummy variables. 

The results are shown in Table \ref{tab:regression}. The log-transformation of the dependent variable complicates the interpretation of the estimated coefficients. Therefore, we calculated marginal effects for each QID scenario and then retransformed them to the original scale~\cite{Duan1983}.

\begin{sidewaystable}
\centering
\small
\def\sym#1{\ifmmode^{#1}\else\(^{#1}\)\fi}
\begin{tabular}{lcccccc}
    \toprule
    &\multicolumn{6}{c}{QID set}\\
    \cmidrule{2-7}
    &\multicolumn{2}{c}{all}&\multicolumn{2}{c}{core+PoB}&\multicolumn{2}{c}{core}\\
    \cmidrule(lr){2-3}\cmidrule(lr){4-5}\cmidrule(lr){6-7}
    & coefficient & se & coefficient & se & coefficient & se\\
    \midrule
    $\beta_0$ & 12.98\sym{**} & (0.0528) & 13.05\sym{**}& (0.0498) & 14.44\sym{**} & (0.0409) \\
    error factor\\
    ~~4 \\
    ~~8  & 0.581\sym{**} & (0.0373) & 0.682\sym{**} & (0.0352) & 0.520\sym{**} & (0.0289)\\
    ~~16 & 1.358\sym{**} & (0.0373) & 1.434\sym{**} & (0.0352) & 0.991\sym{**} & (0.0289) \\
    overlap\\
    ~~25\% & -0.883\sym{**} & (0.0373) & -0.883\sym{**} & (0.0352) & -0.852\sym{**} & (0.0289) \\
    ~~50\%\\
    ~~75\% &  0.512\sym{**} & (0.0373) &  0.511\sym{**} & (0.0352) &  0.497\sym{**} & (0.0289) \\
    UID fill\\
    ~~0\% \\
    ~~70\% & -0.748\sym{**} & (0.0373) & -0.810\sym{**} & (0.0352) & -0.886\sym{**} & (0.0289) \\
    ~~90\% & -1.466\sym{**} & (0.0373) & -1.588\sym{**} & (0.0352) & -1.754\sym{**} & (0.0289)\\
    address outdating\\
    ~~2\% & -0.0735\sym{*} & (0.0373) & 0.00106 & (0.0352) & 0.00209 & (0.0289) \\
    ~~4\% \\
    ~~6\% & 0.0843\sym{*}& (0.0373) & 0.0129 & (0.0352) & -0.0000548 & (0.0289) \\
    update probability\\
    
    ~~0\%  &  0.480\sym{**} & (0.0373) &  0.294\sym{**} & (0.0352) &  0.322\sym{**} & (0.0289) \\
    ~~50\% & -0.0412        & (0.0373) & -0.0759\sym{*} & (0.0352) & -0.0368        & (0.0289)\\
    ~~100\% \\
    duplicates &  0.443\sym{**} & (0.0305) &  0.371\sym{**} & (0.0288) &  0.124\sym{**} & (0.0236) \\
    \midrule
    n & 486 && 486 && 486 \\
    $R^2$ & 0.910 && 0.922 && 0.939 \\
    \bottomrule
\multicolumn{4}{l}{\footnotesize \sym{*} \(p<0.05\), \sym{**} \(p<0.01\)}\\
\end{tabular}
\caption{Linear regression models to predict the log-transformed number of false negatives for each QID set.}
\label{tab:regression}
\normalsize
\end{sidewaystable}

\clearpage

\begin{figure}[h]
    \centering
    \includegraphics[width=\textwidth]{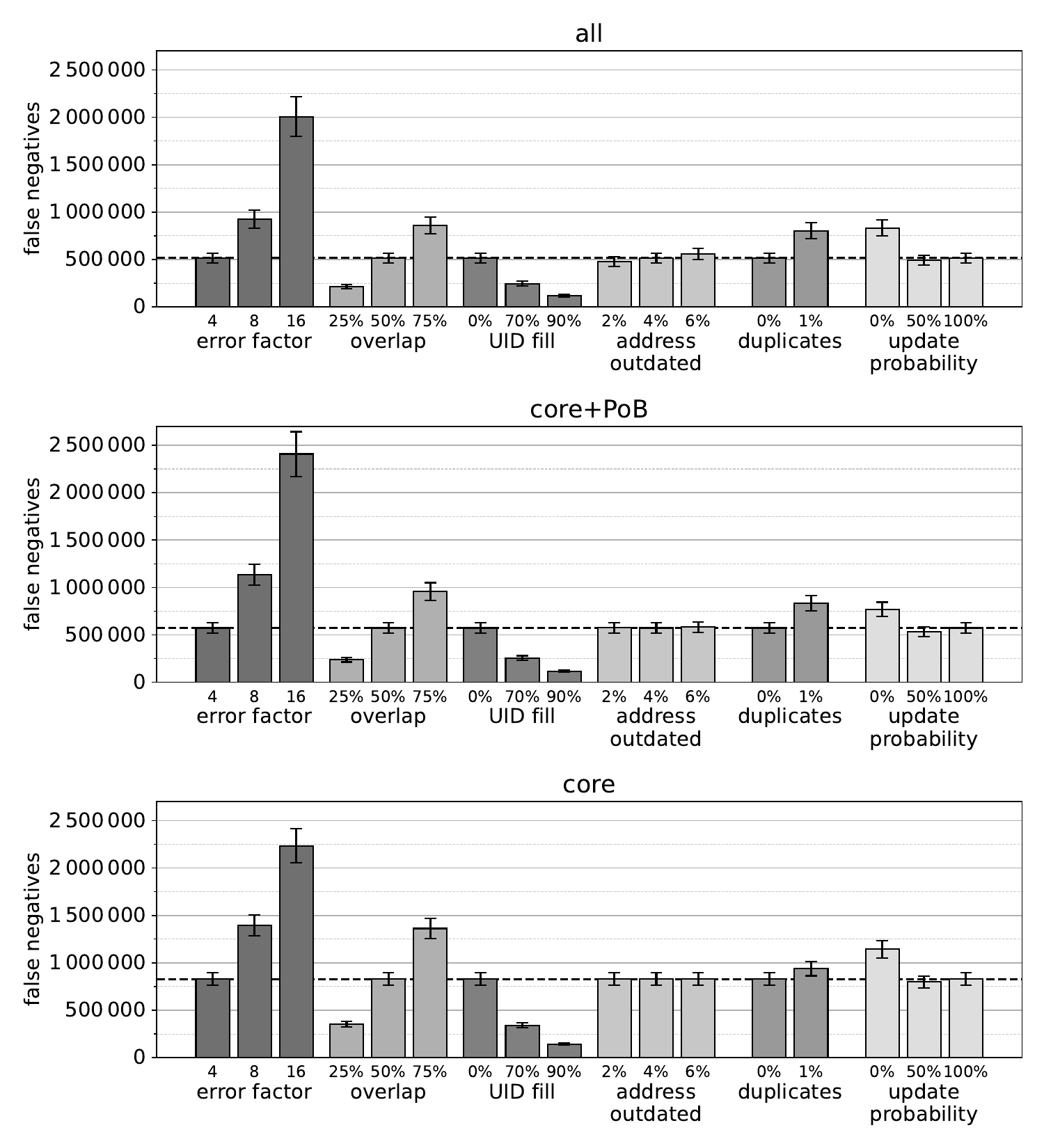}
    \caption{Retransformed marginal effects of the separate linear regression models for each QID set. The dashed lines indicate the predicted outcome at the selected covariate values (error factor 4, 50\% overlap, 0\% UID fill, 4\% address outdating, 0\% duplicates and 100\% update probability).}
    \label{fig:margins}
\end{figure}

Figure \ref{fig:margins} show the marginal effect at error factor 4, 50\% overlap, 0\% UID fill, 4\% address outdating, 0\% duplicates and 100\% update probability for the three QID sets. The results can be interpreted as the estimated change in the number of false negatives when changing one experimental factor while keeping all others constant. 

All experimental factors impacted the number of estimated negatives. The effects of UID fill, overlap and the error factor seem to be the most important predictors of linkage quality. 

\section{Discussion}

The results indicate that the error factor has the highest impact on the number of false negatives. The strength of this factor, as well as most other factors, also depends on the number of available QIDs. In line with expectations, the best results occur when all QIDs are available. However, the core identifiers with place of birth fall slightly behind, indicating that the additional QIDs provide little information gain. With only the core identifiers, the number of false negatives increases by a larger margin. This indicates that the data quality of the registers, in terms of accuracy (error factor) and completeness (available QID set)~\cite{HerzogScheurenWinkler2007}, is of key importance for the register-based census~\cite{Gleich2025}.

Since more records can be subject to error, the number of false positives increases with overlap size. A similar effect can be observed for the UID fill. Only in the most optimistic scenarios does a 90\% UID fill yield fewer than 100,000 false negatives (as indicated in Table~\ref{tab:FPFN}).

The presence of duplicates also raises the number of false positives. This is mainly because a one-to-one linkage constraint can no longer be assumed. This highlights the importance of these constraints and the requirement of deduplicated registers. In particular,  proper deduplication of the resident registers is critical so that at least a one-to-many constraint can be assumed. Otherwise, linkage quality can be much lower.

It is also important to note that no coverage errors in the comparison registers were modelled. Instead, a perfect subset of the population data was assumed. Coverage errors will likely further increase the number of linkage errors, as these records have no true links in the dataset. However, these were not modelled, as little information about the coverage errors of the comparison registers is published. 

The regression results and marginal effects indicate almost no significant impact of the address outdating probability. This is probably because the same net coverage errors were simulated regardless of the outdating probability.

The update probability shows almost no significant difference between 100\% and 50\%, where a 50\% update probability even shows slightly better linkage results. This may be an artefact of the experimental design when sampling records without a UID. Since missing updates are also counted as errors, the number of records with update errors (mainly address errors) increases, resulting in many records having one error when the update probability is 50\%~\cite[p. 54]{WeiandSchnell2026}. However, the linkage method is more robust to these update errors. Because records with errors are prioritised, more of these easier-to-link records are sampled with a 50\% update probability than with a 100\% update probability, resulting in fewer false negatives. An expected increase in false negatives at 50\% update probability can be observed in scenarios with 0\% UID fill.

The findings suggest that for larger registers ($\geq 50\%$ overlap), between 500\,000 and 1\,000\,000 false negatives can be expected in the most optimistic scenarios (low error rate, high update probability, no duplicates), when all QIDs are available. The entropy of the QIDs is therefore not high enough to allow for an error-free linkage of two datasets at this size.\footnote{For the use of entropy in assessing quasi-identifier quality, see~\cite[p. 102]{ChristenRanbadugeSchnell2020}.} This is due not only to QID errors but also to the homogeneity of QIDs within the population, as modelled in the microsimulation framework. Certain QIDs combinations are therefore more probable, making it harder to identify records with highly similar QIDs correctly. Although this is a problem for linkages at all scales, it is especially persistent for large-scale linkages because the probability that a similar record is included in the other dataset is high.

A lower number of false negatives can be expected when a high number of records can be linked with a UID. However, it remains unclear whether the UID will be present in all registers at the time of the census or to what extent these identifiers will be accurate, as no relevant information on this aspect of data quality has been published to date. Assuming the data quality of the tax register is no better than that of the resident registers, we can expect a much smaller reduction in false negatives.

Unresolved cases, such as potential links with multiple candidate pairs or suspected census-relevant false negatives, must be processed through manual or semi-manual clerical edits. As mentioned above (see Table \ref{tab:FPFN}), the highest number of false negatives among the best-case scenarios is 731\,874. Even a medium error rate would result in 450\,630 unclear candidates.\footnote{This is a conservative assumption. The number of candidate pairs is not equal to the number of false negatives; it is substantially higher because it includes additional non-matching pairs.} About 450\,000 cases imply either the use of postal enquiries, which would correspond to a mail survey with the same number of cases, or a manual evaluation of the given information. A conservative estimate for a mail survey is approximately €15 per case, resulting in costs exceeding €6.75 million. Since additional mailings are required for units which cannot be reached by mail or do not respond, the final costs will be substantially higher. If no survey is undertaken and only clerical edits are used, costs may be lower. Depending on the kind of clerical edits, the time needed varies. If more than a glance at QIDs is required, a lower bound of 15 minutes per case would require more than 112\,500 hours, meaning about 66 people working for a year, which would amount to about €3.63 million at current pay rates. Detailed enquiries, or lower data quality, will sharply increase the required money and time. It should be noted that this is the amount of work required for one register. Using a couple of registers within a sign-of-life approach will increase the effort correspondingly.

\section{Limitations}
The results of all simulations depend on the underlying assumptions and data used~\cite{RobinsonBrooks2025}. This is particularly relevant to the assumed error distributions of the QIDs. Additional sources of uncertainty arise from potential coverage errors in the comparison registers and missing data among the QIDs. These factors are not accounted for in the simulation model, as, to the best of our knowledge, no published studies have yet investigated them using German administrative data.

Due to Statistical Disclosure Control, the base dataset contains only frequent household compositions and household sizes, since published official statistics group households with six or more persons 
into a single category. 

The name databases used in the simulations are commercial sources and cover the economically active population. However, commercial databases reflect names in everyday use rather than the sometimes considerably longer names, including multiple additional name components, recorded in official administrative registers. Consequently, names were treated as if they had already undergone standardization and cleaning before linkage. Taken together, these limitations in the underlying data may result in an overestimation of data quality and, consequently, an underestimation of the challenges of national-level record linkage. Both effects could only be assessed through analyses based on administrative data holdings at the federal level. No such analyses have been published by any federal authority.

\section{Conclusion}

Since the tax identification number is not expected to be available in all registers, at least some registers or subpopulations will need to be linked using quasi-identifiers. The simulation results suggest that, under these conditions, errors in QIDs may prevent the rate of false-negative links from falling to an acceptable level. In addition, the entropy of the available QIDs appears to be insufficient to uniquely identify all individuals in a census dataset of this size. Taken together, these findings indicate that it may not be possible to correctly link a non-negligible proportion of the population using the information currently available in the registers. 

Following the legal challenges to the 2011 Census results, which culminated in proceedings before the Federal Constitutional Court, and the current legal proceedings brought by more than 500 municipalities contesting the 2022 Census results, further legal challenges to the 2031 Census results remain a distinct possibility. 

Therefore, given the substantial change in census methodology and the serious challenges to data quality, the next census will require more detailed documentation than the two previous censuses. The United Nations emphasises the importance of a comprehensive quality assurance framework that includes systematic quality assessment as well as transparent documentation and evaluation of census results~\cite{UNSD2022}. However, the current draft bill on implementing the register-based census does not appear to establish any quality measures. To ensure compliance with the principles outlined in the Code of Practice for European Statistics~\cite{Eurostat2018}, the legislation should therefore provide for systematic quality assessment and documentation of potential sources of error.

The magnitude of the false negatives reported here agrees with the estimates resulting from the weapon register and the results reported by~\cite{WojcikSoellner2025}. Therefore, the magnitude of the problem seems to be established now.

The principal limitation of the 2031 Census is the lack of identifying information and substantive data in the administrative registers. Under these conditions, research aimed at developing more efficient AI and machine-learning algorithms for linking the 2031 Census may face considerable challenges. A different census model~\cite{Gleich2025} may therefore offer a more pragmatic solution than accepting multi-year delays and allocating substantial resources to the manual post-processing of millions of potential links.

\subsection*{Statement on Data and Source Code Availability}
The source code of the data generation will be made available in a separate, forthcoming publication. Links to all publicly available data sources are provided in the documentation describing the construction of the base dataset.~\cite{WeiandSchnell2026} Unpublished, commercial, or restricted official statistical datasets used in the simulation cannot be shared due to legal agreements.

\subsection*{Acknowledgements}
We thank Manfred Antoni (IAB) for his assistance in analysing the ALWA data, and Christian Borgs and Virginia Wedekind, former members of the research group, for their support in preparatory work on population simulation. Grammarly was used to support proofreading and language editing, including checks for grammar, spelling, and style.

\subsection*{Funding and Competing Interests}

The authors received no external funding for this research and declare no competing interests.

\bibliography{references}

\end{document}